\documentclass[12pt]{article}
\usepackage{amsmath}
\usepackage{amssymb}
\begin{document}
\begin{flushright}
KFKI-1985-97\\
HU ISSN 0368 5330\\
November 1985
\end{flushright}
\vskip 3truecm
\begin{center}
STOCHASTIC PURE STATE REPRESENTATION\\ FOR OPEN QUANTUM SYSTEMS
\vskip 1truecm
L. Di\'osi
\vskip .5truecm
Central Research Institute for Physics\\
H-1525 Budapest 114, P.O.B. 49, Hungary
\end{center}
\vskip 1truecm
ABSTRACT

We show that the usual master equation formalism of Markovian open
quantum systems  is completely equivalent to a certain state vector formalism.
The state vector of the system satisfies a given frictional  Schr\"odinger 
equation except for random instant transitions of discrete nature. Hasse's
frictional Hamiltonian is recovered for the damped harmonic oscillator.
\pagenumbering{gobble}
\newpage
\pagenumbering{arabic}
The incorporation of damping into the dynamics of quantized
open systems is not a trivial task [1]. Actually, the master
equation formalism operating with mixed states is being considered
as the adequate apparatus in describing damped systems [2,3].

There have been, however, attempts to keep the pure state
representation with a certain nonlinear effective Schr\"odinger 
equation governing the state vector. For example, Hasse [4] found
special solutions of the master equation for the harmonic
oscillator with weak damping.

In our paper we propose a general pure state representation
for Markovian open quantum systems which is completely equivalent
to the density operator formalism. Hasse's result [4] will be
recovered as a special case.

In the mixed state formalism, the evolution of a given
Markovian open system is characterized by the master
equation: 
$$
\dot\rho(t)=L[\rho(t)]
\eqno{(1)}
$$
where $\rho(t)$ is the density operator of the system and $L[\rho]$ is an
operator valued linear function of $\rho$. Since $\rho(t)$ must be normalized
and hermitian, $L$ must be traceless and hermitian as well.

For a given pure state $\psi\psi^\dagger$, where $\psi$ stands for the state
vector, it is useful to introduce the total decay rate $w_{\psi\rightarrow}$ by
$$
w_{\psi\rightarrow}\equiv-\psi^\dagger L[\psi\psi^\dagger]\psi
\equiv -\langle  L[\psi\psi^\dagger]\rangle~~.
\eqno{(2)}
$$
Furthermore, transition rates $w_{\psi\rightarrow\varphi}$ from $\psi$ to another orthogonal
state $\varphi$ can be defined by
$$
w_{\psi\rightarrow\varphi}\equiv-\varphi^\dagger L[\psi\psi^\dagger]\varphi~~,
~~~~\varphi^\dagger\psi=0~~.
\eqno{(3)}
$$
\newpage

If a given set $\{\psi,\varphi_n; n=1,2,\dots,N \}$, $N\leq\infty$, forms  a complete
orthonormal system of state vectors then the identity $\mathrm{tr}L[\psi\psi^\dagger]=0$
leads to the following relation:
$$
\sum_{n=1}^{N} w_{\psi\rightarrow\varphi_n} = w_{\psi\rightarrow}~~, 
\eqno{(4)}
$$
i.e. the partial transition rates to the $\varphi_n$'s sum up to the total
decay rate of the state $\psi$.

Now we turn to develop the pure state equations equivalent
to eq. (1).

Let us introduce the frictional  Schr\"odinger equation for the 
state vector $\psi(t)$:
$$
\dot\psi=(L[\psi\psi^\dagger]-\langle L[\psi\psi^\dagger]\rangle)\psi
\equiv-\frac{i}{\hbar}H_{fr}\psi~~.
\eqno{(5)}
$$
Note that the frictional (non-hermitian and non-linear but norm
conserving) Hamiltonian $H_{fr}$ is not fixed uniquely by the above
formula.

Let us assume now that at $t=0$ the given open system is
prepared in a certain pure state $\rho(0)=\psi(0)\psi^\dagger(0)$. By an infinitesimal
time $dt\equiv\epsilon$ later the state of the system turns out to be
the mixed state
$$
\rho(\epsilon)=\psi(0)\psi^\dagger(0)+\epsilon L[\psi(0)\psi^\dagger(0)]
\eqno{(6)}
$$
according to the master eq. (1). If the state $\psi(t)$ of the system
satisfied the  frictional  Schr\"odinger equation (5) then the
density operator  $\rho(\epsilon)$ would be equal to
$$
\psi(\epsilon)\psi^\dagger(\epsilon)
=\psi(0)\psi^\dagger(0)-\frac{i\epsilon}{\hbar}[H_{fr}\psi(0)\psi^\dagger(0)-\psi(0)\psi^\dagger(0)H_{fr}]=~~~~~~~~~~
\eqno(7)
$$
$$
~~~~~~~=\psi(0)\psi^\dagger(0)+\epsilon\{L[\psi(0)\psi^\dagger(0)]-\langle L[\psi(0)\psi^\dagger(0)]\rangle,\psi(0)\psi^\dagger(0)\}
$$
which is a pure state of course. Now, the true state (6) can be
formally written as
$$
\rho(\epsilon)=\psi(\epsilon)\psi^\dagger(\epsilon)+\epsilon W
\eqno{(8)}
$$
\newpage
{\noindent if we introduce the hermitian operator $W$ as the following function
of the actual state $\psi$:}
$$
W = L - \{L-\langle L\rangle,\psi\psi^\dagger\}
\eqno(9)
$$
where the $\psi$-dependence of $L\equiv L[\psi\psi^\dagger]$ is understood. We shall call
$W$ the \emph{transition rate operator} in the given state $\psi$ of the system.
Using definition (9), it is trivial to show that the transition
rates (3) can equally be expressed by $W$, too:
$$
w_{\psi\rightarrow\varphi}\equiv \varphi^\dagger L\varphi = \varphi^\dagger W\varphi~~,~~~~\varphi^\dagger\psi=0~~;
\eqno{(10)}
$$
$$
w_{\psi\rightarrow}\equiv -\langle L\rangle = \langle W\rangle~~.~~~~~~~~~~~~~~~~~~~~
$$

Observe that $\psi$ itself is the eigenvector of $W$ with the 
eigenvalue $-w_{\psi\rightarrow}$:
$$
W\psi = (L-\{L-\langle L\rangle,\psi\psi^\dagger\})\psi = \langle L\rangle\psi = -w_{\psi\rightarrow}\psi~~.
\eqno{(11)}
$$
This genuine property of the transition rate operator $W$ is
strongly correlated with the actual choice (5) of the frictional 
Hamiltonian equation of motion.

Let us suppose now that the transition rate operator (9) has
always a discrete spectrum. If $\{\psi,\varphi_n; n=1,2,\dots,N \}$, $N\leq\infty$,  is the
complete orthonormal system of the eigenvectors of $W$ then we obtain
the following orthogonal expansion:
$$
W = -w_{\psi\rightarrow}\psi\psi^\dagger + \sum_{n=1}^N w_{\psi\rightarrow\varphi_n}\varphi_n\varphi^\dagger_n
\eqno(12)
$$
where the eigenvalues were replaced by the corresponding transition
rates (10).

Let us substitute such an orthogonal expansion into the rhs
of the eq. (8):
$$
\rho(\epsilon)=(1-\epsilon w_{\psi(0)\rightarrow})\psi(\epsilon)\psi^\dagger(\epsilon)
+\epsilon \sum_{n=1}^N w_{\psi(0)\rightarrow\varphi_n(0)}\varphi_n(0)\varphi_n^\dagger(0)~~.
\eqno(13)
$$
\newpage
{\noindent And now, we propose the following statistical interpretation of
the above orthogonal expansion of the mixed state $\rho(\epsilon)$: Given the
pure state $\psi(0)$ at $t=0$, for the infinitesimal period $dt=\epsilon$ the
state vector $\psi(t)$ satisfies the frictional  Schr\"odinger equation
(5) with probability $(1-\epsilon w_{\psi\rightarrow})$ but, alternatively, $\psi(t)$ can decay
into a given eigenvector $\varphi_n$ ot the actual transition rate operator
(9), with the transition rate $w_{\psi\rightarrow\varphi_n}=\varphi^\dagger_n W\varphi_n$ (10), for
$n=1,2,\dots,N$ respectively.}

It is well-known that for a mixed state like $\rho(\epsilon)$ (13) the
statistical interpretation by a certain mixture of pure states is
never unique [5]. Nevertheless, our choice (13) is distinguished
by the mutual orthogonality of the pure states which are the
terms of the decomposition. This decomposition is highly preferred
from the viewpoint [6] of measurement theory, too.

In practice it will turn out that a slightly modified form
$W'$, instead of the transition rate operator $W$, will be more convenient
to use:
$$
W'\equiv L - \{ L,\psi\psi^\dagger\} + \langle L\rangle\psi\psi^\dagger~~.
\eqno(14)
$$
By comparing eq. (14) with eq. (9) one can see that $W'=W-\langle L\rangle\psi\psi^\dagger=$
=$W+w_{\psi\rightarrow\psi\psi^\dagger}$
thus $W'$ will be positive semidefinite with the orthogonal
expansion similar to (12):
$$
W' = \sum_{n=1}^N w_{\psi\rightarrow\varphi_n}\varphi_n\varphi^\dagger_n~~.
\eqno(15)
$$

Recalling the statistical interpretation of the continual
state mixing expressed by eq. (13) we are going to formulate the
stochastic pure state representation.

We propose to characterize the state of the given open
quantum system by the state vector $\psi(t)$, which is, in this case,
to be taken as stochastic variable with vector values from the
Hilbert space of states. The the density operator $\rho(t)$ is
recovered by the stochastic mean of the pure state density
operator $\psi(t)\psi^\dagger(t)$:
\newpage
$$
\rho(t)=\prec\psi(t)\psi^\dagger(t)\succ
\eqno(16)
$$
where the symbol $\prec\succ$ stands for the stochastic mean.

The state vector $\psi(t)$ obeys to the following Markovian
stochastic process: $\psi(t)$ satisfies the continuous and deterministic
Schr\"odinger equation (5) with frictional Hamiltonian
but, at any instant, the state $\psi(t)$ can promptly decay into a
certain eigenvector $\varphi_n(t)$ of the actual transition rate operator
$W'$ (14) with the transition rate $w_{\psi(t)\rightarrow\varphi_n(t)}$ which is equal to
the corresponding eigenvalue of $W'$ (cf. the expansion (15)).

By the construction of this pure state representation (and
especially from eq. (13)) it follows that the density operator
(16) will obey to the original master eq. (1) of the give  system.
Therefore the pure state representation yields identical physical
predictions to the density operator formalism.

Finally, we apply our pure state equations to the damped
harmonic oscillator. This open system is understood relatively
well and the corresponding literature is remarkably wide [1].

The master eq. (1) has now the following form:
$$
\dot\rho=L[\rho]=-\frac{i}{\hbar}[H_0,\rho]-\frac{i}{\hbar}\lambda[x,\{p,\rho\}]
-\frac{1}{\hbar}\mathrm{Re}D_{ab}[A_a,[A_b,\rho]]~~,
\eqno{17}
$$
where $H_0$ is the free oscillator Hamiltonian. For the convenience
of notations, the vector $A$ stands for the pair of the canonical
variables: $A_1=p$ and $A_2=x$. Consequently, $D$ is a given $2\times2$ positive
[3,7] hermitian matrix: $D_{11},D_{22},\mathrm{Re}D_{12}$ are the diffusion
coefficients, $\lambda=\frac{2}{\hbar}\mathrm{Im}D_{12}$ is the constant of friction. On eq. (17)
and henceforth, we use Einstein's convention for summation over
repeated indices.

Applying the formula (5) we are led to the following frictional 
Schr\"odinger equation for the damped harmonic oscillator:
$$
\dot\psi=(L-\langle L\rangle)\psi=~~~~~~~~~~~~~~~~~~~~~~~~~~~~~~~~~~~~~~~~~~~~~~~~~~~~~~~~~~~~
$$
$$
=-\frac{i}{\hbar}(H_0-\langle H_0\rangle)\psi
-\frac{i}{\hbar}\lambda(xp+x\langle p\rangle-\langle x\rangle p-x\langle p\rangle)\psi-~~~~~
\eqno(18)
$$
$$
~~~~~~~~~~~~~~~~~~~~~~~~~~~~~~-\frac{i}{\hbar^2}\mathrm{Re}D_{ab}[(A_a-\langle A_a\rangle)(A_b-\langle A_b\rangle)-\sigma_{ab}]
$$
\newpage
{\noindent where the positive hermitian $2\times2$ matrix $\sigma$ is defined by}
$$
\sigma_{ab}\equiv\langle A_a A_b\rangle-\langle A_a\rangle\langle A_b\rangle~~.
\eqno(18a)
$$
The above equation of motion is obviously consistent with the
following choice for the frictional Hamiltonian $H_{fr}$:
$$
H_{fr}=H_0+\lambda[\frac{1}{2}\{x,p\}-\frac{1}{2}\langle\{x,p\}\rangle+x\langle p\rangle-\langle x\rangle p] -
\eqno(19)
$$
$$
~~~~~~~~~~~~~~-\frac{i}{\hbar}\mathrm{Re}D_{ab}[(A_a-\langle A_a\rangle)(A_b-\langle A_b\rangle)-\sigma_{ab}]~~.
$$

The antihermitian term of this Hamiltonian differs from
Hasse's Hamiltonian [1,4]. Recall, however that Hasse's derivation
requires an additional constraint for the state $\psi(t)$, namely
$$
\mathrm{Re}D_{ab}~\sigma_{ab}=\frac{1}{2}\hbar^2\lambda~~.
\eqno(20)
$$
If this condition is fulfilled our Hamiltonian (19) will be
identical to that of the Hasse theory.

Now, applying the definition (14) to the operator $L$ in the
master eq. (17), we obtain the modified transition rate operator
of the damped oscillator in the following compact form:
$$
W'=\frac{2}{\hbar^2} D_{ab}(A_a-\langle A_a\rangle)\psi\psi^\dagger(A_b-\langle A_b\rangle)
\eqno(21)
$$
which is explicitly positive semidefinite. The rank of $W'$ is
two thus its decomposition (15) contains only two states $\varphi_1$ and
$\varphi_2$ which are orthogonal to each other and to the state $\psi$ of the
system:
$$
W'=w_{\psi\rightarrow\varphi_1}\varphi_1\varphi_1^\dagger
     +w_{\psi\rightarrow\varphi_2}\varphi_2\varphi_2^\dagger
\eqno(22)
$$

The total decay rate of the actual state $\psi$ is given by the
trace of the modified transition rate operator $W'$:
$$
w_{\psi\rightarrow}=w_{\psi\rightarrow\varphi_1}+w_{\psi\rightarrow\varphi_2}
                                      =\frac{2}{\hbar^2}D_{ab}\sigma_{ab}
                                      =\frac{2}{\hbar^2}\mathrm{Re}D_{ab}-\lambda~~.
\eqno(23)
$$
\newpage

It is interesting to recognize that $w_{\psi\rightarrow}$ vanishes if and only 
if the Hasse's condition (20) fulfills for the state $\psi$ of the system.
For such states there is no stochastic transitions and thus 
the frictional Schr\"odinger equation (18) is satisfied exactly,
until the condition (20) is fulfilled, of course.

In general, the diagonalization (22) of $W'$ is trivial, and
the eigenvectors are of the form $\varphi_a=C_{ar}(A_r-\langle A_r\rangle)\psi$ with coefficients $C$ depending on the matrices $D$ and $\sigma$.

For brevity, however, we shall consider the simplest but
still interesting damped oscillator where $D_{11}=D_{12}=D_{22}=0$, $D_{22}>0$.
The frictional Hamiltonian (19) is then the following:
$$
H_{fr}=H_0-\frac{i}{\hbar}D_{22}[(x-\langle x\rangle)^2-\sigma_{22}]~~.
\eqno(24)  
$$

The transition rate operator (21) is now degenerate and
equals to a single diad: 
$W'=2\hbar^{-2}D_{22}(x-\langle x\rangle)\psi\psi^\dagger(x-\langle x\rangle)
       =w_{\psi\rightarrow\varphi}\varphi\varphi^\dagger$
where 
$$
\varphi=(\sigma_{22}^{-1/2})(x-\langle x\rangle)\psi
\eqno(25)
$$       
is normalized and orthogonal to $\psi$. The corresponding transition
rate is then equal to
$$
w_{\psi\rightarrow\varphi}=\frac{2}{\hbar^2}D_{22}\sigma_{22}~~.
\eqno(26)
$$   

Now, we claimed that, on one hand, the state vector $\psi(t)$
satisfies the Schr\"odinger equation with the frictional Hamiltonian
(24); on the other hand, $\psi(t)$ can jump stochastically into the
state $\varphi(t)$ (25) according to the time-dependent transition rate
(26). We are going to prove that the stochastic mean 
$\rho(t)=\prec\psi(t)\psi^\dagger(t)\succ$ (16) 
satisfies the corresponding master equation (17).

If we suppose $\psi(0)\equiv\psi$ is fixed then, by $dt\equiv\epsilon$ later, the state\linebreak
$\psi(t)$ equals to $\psi-\frac{i\epsilon}{\hbar}H_{fr}\psi$, 
with probability $1-\epsilon w_{\psi\rightarrow\varphi}$ and,
alternatively, $\psi(\epsilon)=(\sigma_{22}^{-1/2})(x-\langle x\rangle)\psi=\varphi$,
with probability $\epsilon w_{\psi\rightarrow\varphi}$. 
Therefore the density operator is as follows:
\newpage
$$
\rho(\epsilon)\vert_{\rho(0)=\psi\psi^\dagger}
=(1-\epsilon w_{\psi\rightarrow\varphi})
  (\psi-\frac{i\epsilon}{\hbar}H_{fr}\psi)(\psi^\dagger+\frac{i\epsilon}{\hbar}\psi^\dagger H_{fr}^\dagger)
 +\epsilon w_{\psi\rightarrow\varphi}\varphi\varphi^\dagger~~.
\eqno(27)
$$

From the eqs. (24-26) we can substitute $H_{fr},\varphi$ and $w_{\psi\rightarrow\varphi}$,
respectively, on the rhs of eq. (27). Recalling that $\epsilon$ is infinitesimal,
eq. (27) can be written as
$$
\rho(\epsilon)\vert_{\rho(0)=\psi\psi^\dagger}=\psi\psi^\dagger-\frac{i\epsilon}{\hbar}[H_0,\psi\psi^\dagger]
                                                      -\frac{\epsilon}{\hbar^2}D_{22}[x,[x,\psi\psi^\dagger]~~.
\eqno(28)
$$
Remember this is the density operator at $t=\epsilon$, with the condition
that $\rho(0)$ was a given pure state $\psi\psi^\dagger$. It is easy to remove this
condition because the rhs of eq. (28) is linear in $\psi\psi^\dagger$. By averaging
for all stochastically possible states $\psi(0)$ we can replace 
$\psi\psi^\dagger$ by $\rho(0)$ and we see that eq. (28) will be equivalent to the
corresponding master eq. (17):
$$
\rho(\epsilon)=\rho(0)-\frac{i\epsilon}{\hbar}[H_0,\rho(0)]
                                                      -\frac{\epsilon}{\hbar^2}D_{22}[x,[x,\rho(0)]~~.
\eqno(29)
$$

Thus, for the simple damped harmonic oscillator we have
directly shown the equivalence of the master equation formalism
and the stochastic pure state formalism proposed in the present
paper. Note the first simple example was given in our previous 
work [8] on systems with white-noise potentials. The proposal
has now been generalized and extended for all Markovian open
quantum systems.

We should admit, however, that we do  not know almost
anything about the characteristic solutions of the stochastic
pure state equations except for their statistical equivalence
with the density operator. This question needs further investigations.

I wish to thank to Dr. P. Hrask\'o for the illuminating discussions
and remarks.
\newpage

\parskip 0truecm
\vskip 0.5 truecm
\noindent
REFERENCES
\vskip .2truecm

\noindent\hskip 10pt [1] H. Dekker, Phys. Rep. \underline{80} (1981) 1
\vskip .2 truecm

\noindent\hskip 10pt [2] V. Gorini et al., Rep. Math. Phys.  \underline{13} (1978) 149
\vskip .2 truecm

\noindent\hskip 10pt [3] G. Lindblad, Commun. Math. Phys.  \underline{48} (1976) 119
\vskip .2 truecm

\noindent\hskip 10pt [4] R.W. Hasse, Phys. Lett. \underline{85B} (1979) 197
\vskip .2 truecm

\noindent\hskip 10pt [5] A. Einstein, B. Podolsky and N. Rosen,  Phys. Rev. \underline{47} (1935) 777
\vskip .2 truecm

\noindent\hskip 10pt [6] L. Di\'osi, to be published
\vskip .2 truecm

\noindent\hskip 10pt [7] H. Dekker and M.C. Valsakumar, Phys. Lett. \underline{104A} (1984) 67
\vskip .2 truecm

\noindent\hskip 10pt [7] L. Di\'osi, Phys. Lett. \underline{112 A} 288 (1985)
\vskip .2 truecm

\end{document}